\documentclass[a4paper, aps, prl, longbibliography, nofootinbib,twocolumn,superscriptaddress]{revtex4-1}

\usepackage{asymptote} 
\usepackage{array}
\usepackage{amsmath,amssymb,amsfonts}
\usepackage[colorlinks]{hyperref}
\usepackage{multirow}
\usepackage{float} 
\usepackage[utf8]{inputenc} 
\usepackage{graphicx}
\usepackage{subfigure}
\graphicspath{{./pic/}}

\usepackage[utf8]{inputenc}
\newcommand{\bw}{\begin{widetext}}
\newcommand{\ew}{\end{widetext}}
\newcommand{\be}{\begin{equation}}
\newcommand{\en}{\end{equation}}
\newcommand{\bee}{\begin{equation}}
\newcommand{\ene}{\end{equation}}
\newcommand{\bea}{\begin{eqnarray}}
\newcommand{\ena}{\end{eqnarray}}
\newcommand{\bes}{\begin{subequations}}
\newcommand{\ens}{\end{subequations}}
\newcommand{\bef}{\begin{figure}}
\newcommand{\enf}{\end{figure}}

\def\etc{{\it etc.~}}

\def\calb{\mathcal{B}}

\def\cald{\mathcal{D}}

\def\calj{\mathcal{J}}
\def\calk{\mathcal{K}}
\def\call{\mathcal{L}}

\def\calp{\mathcal{P}}

\def\to{\rightarrow}

\def\d{{\rm d}}

\def\gev{{\rm GeV}}

\begin{document}


\title{Kinematical Observables in Semi-Invisible Decays}

\author{Kai Ma}
\email[Electronic address: ]{makai@ucas.ac.cn}
\affiliation{School of Fundamental Physics and Mathematical Science, Hangzhou Institute for Advanced Study, UCAS, Hangzhou 310024, Zhejiang, China}
\affiliation{International Centre for Theoretical Physics Asia-Pacific, Beijing/Hangzhou, China}
\affiliation{Department of Physics, Shaanxi University of Technology, Hanzhong 723000, Shaanxi, China}

\date{\today}

\begin{abstract}
Invisible particles frequently appear in final state in studying physics at colliders.
Experimental precision is also low in measuring missing energy. In this paper,  
we propose a general approach for studying process involving invisible particles.
We provided two kinematical observables which are sensitive to production kinematics in different regions, and hence are complementary. Usage of our approach is illustrated by
three examples. It is shown that our observables are still useful in case of that 
the significance $S/B$ is relatively low.
\end{abstract}

\maketitle


%

Observation of the Higgs scalar $h(125)$~\cite{Aad:2012tfa,Chatrchyan:2012ufa} 
implies the final building block
of the Standard Model (SM) was found. However, it does not mean one have 
completely understand our universe~\cite{10.1093/ptep/ptaa104}. 
On the one hand, physical properties
of the mass eigenstate  $h(125)$ are still not measured at a requisite
precision~\cite{Aad:2019mbh,Sirunyan:2018koj}, 
for instances $CP$ nature, potential form \etc. 
On the other hand,
cosmological and astronomical observations convincingly indicate that 
matter particles predicted by the SM are only a few percent of our universe,
new particles with a broad range of mass have to exist 
in~\cite{Simon:2019ar,Salucci:2018hqu}. In general,
particles beyond the SM are neutral with respect to charges in the 
SM~\cite{McDermott:2010pa},
and hence weakly interacting with the particles that have been 
observed~\cite{Randall:2007ph,Tulin:2017ara}.
Therefore, invisible they are usually in experimental detectors, and appear 
as missing energy of which the experimental precision is low. 

This is not only a general property in beyond standard model (BSM),
the neutrinos as the lightest sector in the SM are also invisible at current
colliders. Experimental investigations on processes involving neutrinos 
turn out to be rather hard, particularly at a hadron collider~\cite{Webber:2009vm}. 
The situation
becomes worse when the collider energy is increased, where 
electroweak radiations and hadronization effects can further reduce 
measurement precision on the missing energy~\cite{Nojiri:2010mk}.

However, experimental searches involving those invisible particles are 
particularly important~\cite{ATLAS:2018iyk}. 
For instance, the lightest neutrilino unavoidably 
appear in hunting for supersymmetry particles~\cite{Cho:2007dh};
observation of neutrino oscillations not only proves that neutrinos are massive,
but also implies lepton flavor number violated processes~\cite{Cheng:1977nv,Lee:1977tib,Marciano:1977cj,Marciano:1977wx},
which could give constraints 
to the lepton mixing angles and the neutrino mass ratios, 
and also provide an alternative window for probing properties of new physics.

\begin{figure}[h]
\begin{center}
\includegraphics[scale=0.58]{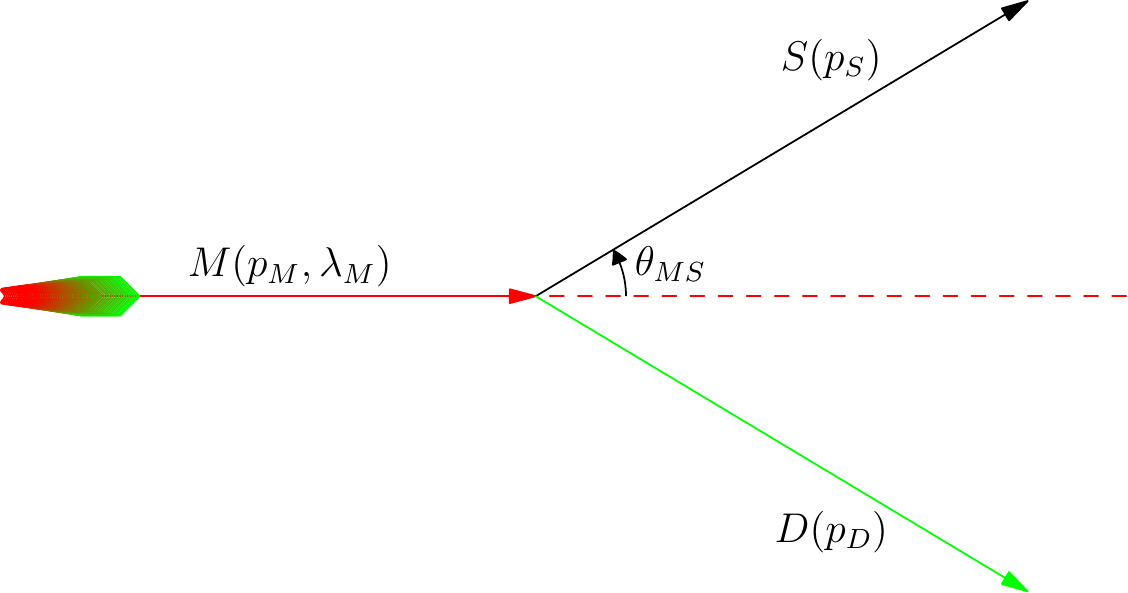}
\caption{Sketch plot of the decay process $M(p_{M}, \lambda_{M}) \to S_{p_{S}} + D(p_{D})$,
where $\lambda_{M}$ accounts for possible helicity values of the mother particle $M$. $S$ is
visible and $D$ is invisible.}
\label{fig:Decay:Conf}
\end{center}
\end{figure}

Therefore, essentially important the frequently happening 
semi-invisible decays are~\cite{HarlandLang:2012gn,HarlandLang:2011ih,Nojiri:2010dk}.
In this letter we report a general method for studying processes involving invisible particles. 
Without loss of generality, we assume that the invisible particle $D$ emerges from
decay of  a mother particle $M$ which is generated
in some certain mechanism. Furthermore, the associated particle $S$ is taken to be visible.
Fig.~\ref{fig:Decay:Conf} is a sketch plot of such a configuration.
Theoretically, mass of the invisible particle $D$ can be uniquely determined
if the rest frame of the mother particle $M$ can be precisely reconstructed.
However, in practice $M$ are produced in pair, and hence it is nearly
impossible to identify the missing momentum. On the other hand,
since the invisibility of $D$, spin correlation effects which are powerful probes 
of production and decay dynamics are covered up. 
Here we introduce a new method to study the semi-invisible decay modes.

Under narrow width approximation, 
the total cross section with subsequent decay 
$M(p_{M}, \lambda_{M}) \to S_{p_{S}} + D(p_{D})$ 
can be written in general as,
\bee
\frac{ \d \sigma }{ \d \varPhi_{M} \d \varPhi_{S}^{\star} }
=
\frac{ 1 }{ 2 m_{M} \Gamma_{M} }
\calp_{\lambda_{M}}^{ \lambda'_{M}}( p_{M} )\,
\cald_{\lambda_{M}}^{ \lambda'_{M}}( p_{M}^{\star}, p^{\star}_{S} ) \,,
\ene
where $\calp_{\lambda_{M}}^{ \lambda'_{M}}( p_{M} )$ and
$\cald_{\lambda_{M}}^{ \lambda'_{M}}( p_{M}^{\star}, p^{\star}_{S} )$
are production and decay helicity density matrix of the mother particle $M$, 
and the helicity values $\lambda_{M}$ are summed over; futhermore
$ p_{M}^{\star}$ and $p^{\star}_{S} $ are momentum of the particle $M$
and $S$ in the rest frame of $M$, respectively; and the decay helicity density
matrix is calculated also in this frame where spin polarization effects are maximum.
However, practically useful differential cross section is represented in terms
of kinematical variables which can be directly measured in experiments.
Then moving into the Lab. frame, the differential cross section is given as
\bee
\frac{ \d \sigma }{  \d \varPhi_{M} \d \varPhi_{S} }
=
\frac{ \calp_{\lambda_{M}}^{ \lambda'_{M}}( p_{M} ) }{ 2 m_{M} \Gamma_{M} }
\,
\cald_{\lambda_{M}}^{ \lambda'_{M}}( p_{M}^{\star}, p^{\star}_{S} ) \,
\calj( p_{M}; p_{S}^{\star}, p_{S} )  \,,
\ene
where $\calj( p_{M}; p_{S}^{\star}, p_{S} )$ is the Jacobi factor which
stands for the variable transformation from $p^{\star}_{S}$ to $p_{S}$.
In consideration of that
$\cald_{\lambda_{M}}^{ \lambda'_{M}}( p_{M}^{\star}, p^{\star}_{S} )$
is un-measurable due to the missing momentum $p_{D}$, the polarization
effects have to be averaged, and then we have,
\bee\label{eq:factorization:verage}
\frac{ \d \sigma }{  \d \varPhi_{M} \d \varPhi_{S} }
=
\frac{ \overline{\calp}( p_{M} ) }{ 2 m_{M} \Gamma_{M} }\,
\overline{\cald}( p_{M}^{\star}, p^{\star}_{S} ) \,
\calj( p_{M}; p_{S}^{\star}, p_{S} )  \,.
\ene
In this form, the production helicity density matrix 
$\calp_{\lambda_{\tau}}^{ \lambda'_{\tau}}(p_{M})$
is completely factorized out. For given production mechanism of the 
mother particle $M$ with momentum $p_{M}$, 
the production helicity density matrix 
$\calp_{\lambda_{\tau}}^{ \lambda'_{\tau}}(p_{M})$
is unique, and hence can provide a template for studying the decay dynamics.
Furthermore, the decay helicity density matrix 
$\cald_{\lambda_{\tau}}^{ \lambda'_{\tau}}$
in which the decay dynamics is completely encoded is also factorized out
in the sense of that it is independent of $p_{M}$. 
On the other hand, the Jacobi factor 
$\calj( p_{M};  p_{S}^{\star}, p_{S}  )$ can be directly calculated once 
$p_{M}$ and $p_{S}$ are given 
($p_{S}^{\star}$ is expressed in terms of $p_{M}$ and $p_{S}$).

Therefore, factorization in Eq.~\eqref{eq:factorization:verage} implies that
the kinematical distributions of the particle $S$ can be investigated 
by integrating over the momenta $p_{S}$ for given $p_{M}$. Based on this 
consideration we introduce following two observables,
\bea
m^{+}_{D}( p_{M} ) 
&=&
\left\{ \sqrt{ \big| \text{Max}\left\{ m_{D}^{2}\right\} \big|}  
 \bigg| \; \overline{\cald}( p^{\star}_{S} ) \,
\calj(p_{S}^{\star}, p_{S} )   \right\} \,,
\\[3mm]
m^{-}_{D}( p_{M} ) 
&=&
\left\{ \sqrt{ \big| \text{Min}\left\{ m_{D}^{2}\right\} \big|}  
 \bigg| \; \overline{\cald}(p^{\star}_{S} ) \,
\calj( p_{S}^{\star}, p_{S} )   \right\} \,,
\ena
where $m_{D}^{2} = (p_{M} - p_{S})^{2}$. 
Then the boundary values are simply given as
\bee
m_{D}^{\pm} 
= 
\sqrt{\left| m_{M}^{2} + m_{S}^{2} - 2 E_{M} E_{S} ( 1  \mp  \beta_{M}  \beta_{S}) \right| } \,.
\ene
Here $E_{S}$ and $\beta_{S}$ are functions of $m_{D}^{2}$, $p_{S}^{\star}$,
as well as $p_{M}$. Neglecting the $p_{M}$ and $p_{S}^{\star}$ dependences 
of $m^{\pm}_{D}$, in the massless limit $m_{S}=0$ we have
\bee\label{eq:mpm:massless}
m_{D}^{\pm} 
= \sqrt{ \left| m_{M}^{2} + \dfrac{ 1 \mp \beta_{M} }{ 1 \pm \beta_{M} } 
\left( m_{D}^{2} - m_{M}^{2} \right) \right| } \,.
\ene
We can see that in this case $m_{D}^{+}$ is maximumly sensitive to 
$m_{D}$ when $\beta_{M} = 0$, while $\beta_{M} = 1$ the most sensitive 
region is for $m_{D}^{-}$. In this sense $m_{D}^{+}$ and $m_{D}^{-}$
are two complementary observables for studying the whole kinematical region.

Differential distribution with respect to the variables $m^{\pm}_{D}$ 
can be written as,
\bee
\frac{ \d\sigma}{  \d m^{+}_{D} \d m^{-}_{D} }
=
\int \d \varPhi_{M}
\frac{ \overline{\calp}( p_{M} ) }{ 2 m_{M} \Gamma_{M} }  \,
\overline{\calb}( p_{M}, m^{\pm}_{D} )\,,
\ene
where the weight function $\overline{\calb}( p_{M}, m^{\pm}_{D} )$ is given as
\bee
\overline{\calb}
=
\int \d \varPhi_{S} \,
\overline{\cald}( p_{S}^{\star} ) \,
\calj(p_{S}^{\star}, p_{S} ) \,
\delta^{2}\left(m^{\pm}_{D} - \calk^{\pm}( p_{M}^{\star}, p_{S}^{\star} ) \right)\,.
\ene
Here $\calk^{\pm}( p_{M}^{\star}, p_{S}^{\star} )$ are kernel functions that give
the boundary values $m^{\pm}_{D}$ by integrating over the kinematical variables
$p_{S}$ for given $p_{M}$. Therefore, the distribution shapes of $m_{D}^{\pm}$
encode the decay dynamics (production dynamics given by $\overline{\calp}( p_{M} )$
is universal). As long as the kernel functions don't spread too much, 
$m_{D}^{\pm}$ have maximum sensitivities to $m_{D}$.
We illustrate usage of our observables with three examples bellow.

\textbf{\em{Example I: Lepton flavor violating decay of the $\tau$-lepton}}.
The $\tau$-lepton, as the heaviest particle in the lepton sector, is expected
to be a natural probe for lepton number violation. The present best limit of 
$\calb(\tau\to3\mu) < 2.1 \times10^{-8}$ at 90\%CL was obtained by the 
Belle experiment~\cite{Hayasaka:2010np}, and a slightly mild upper limit of 
$6.9 \times10^{-8}$ at 90\%CL was obtained by the CMS experiment~\cite{CMS:2020kwy}.
On the other hand, invisible decay modes of the $\tau$-lepton, $\tau\to\ell\alpha$, 
can also be essential~\cite{GRINSTEIN198557,Feng:1997tn,Asai:2018ocx}
when new $Z'$ gauge bosons~\cite{Altmannshofer:2016jzy,Altmannshofer:2016brv,Heeck:2016xkh} 
or axion-like particles~\cite{PhysRevLett.49.1549,Berezhiani:1989fp,Calibbi:2020jvd} exist in nature.
Because of large irreducible backgrounds from the leptonic decay modes 
$\tau \to \ell \nu_{\tau} \bar{\nu}_{\ell}$,
experimental constrains on these decay modes are relatively weak.
The present upper limits of the branching fractions are
obtained by the ARGUS collaboration~\cite{Albrecht:1995ht},
and are several percent of the corresponding leptonic decay modes.
The upper limits in Ref.~\cite{Albrecht:1995ht} are obtained by tagging 
one side of the $\tau$-lepton pair decays in 3-prong mode.
A new method was proposed in Ref.~\cite{DeLaCruz-Burelo:2020ozf}. 

\begin{figure}[b]
\begin{center}
\includegraphics[scale=0.238]{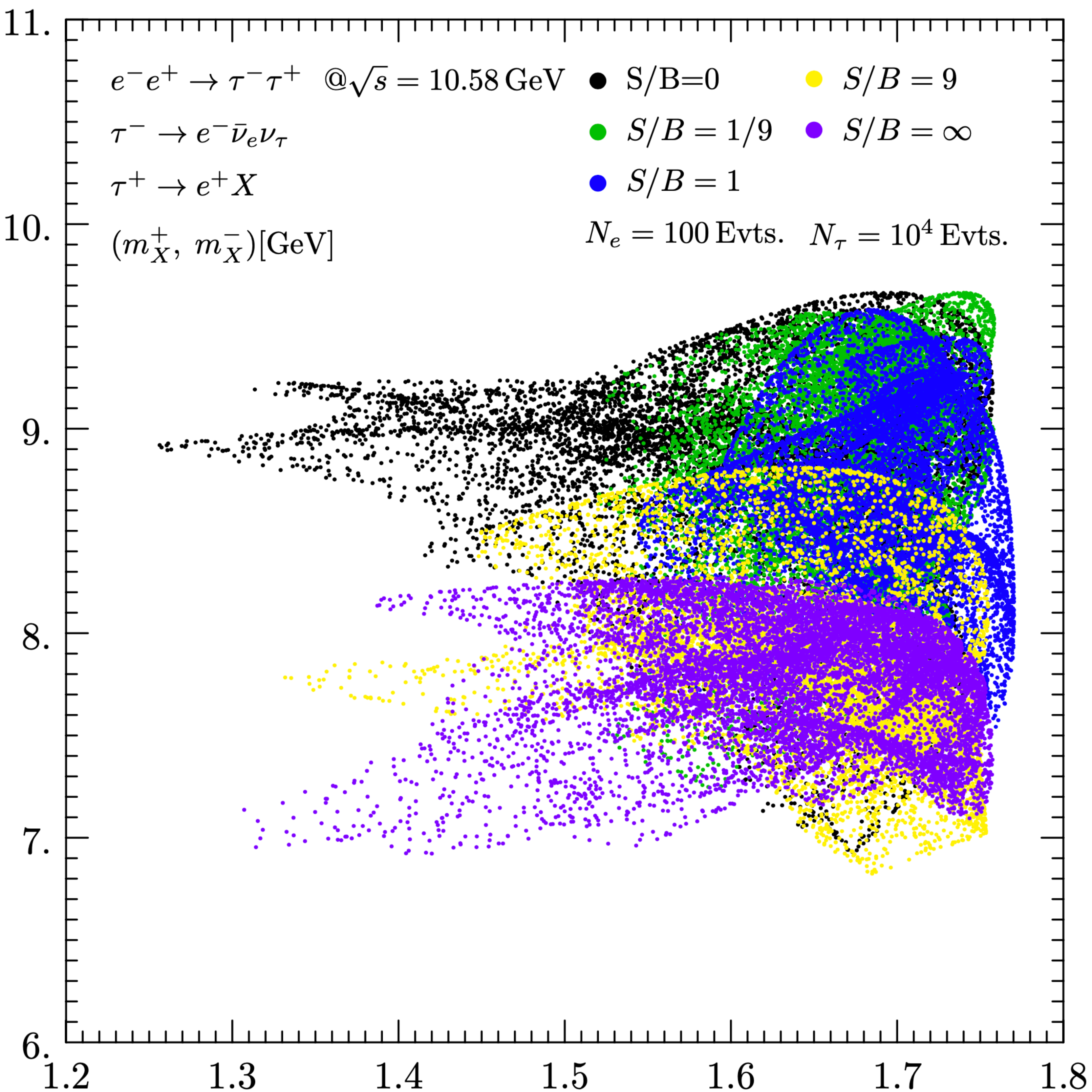}
\caption{scatter plots in the $(m_{D}^{+},\, m_{D}^{-})$ plane for
 different significance $S/B$ with $100$ total number of events and $10^{4}$ events
for the templates of $\tau^{\pm}$.
}
\label{fig:MassX:Mix}
\end{center}
\end{figure}

Here we study how our observables behaves for the decay process $\tau\to\mu X$
at the energies of the Belle II 
experiment~\cite{Abe:2010gxa,Kou:2018nap,Villanueva:2018pbk,Konno:2020+A}. 
The interaction Lagrangian is simply given a usual Yukawa interaction, 
$
\call_{I} = -g_{X} X \overline{\tau} \mu + h.c.
$
and was implemented in FeynRules~\cite{Alloul:2013bka}. 
Events are generated at $\sqrt{s}=10.58\gev$ 
by using MadGraph~\cite{Alwall:2014hca}.
\begin{widetext}
\mbox{}
\begin{figure}[h]
\begin{center}
\subfigure[]{\label{fig:MassX:Max}}
{\includegraphics[scale=0.578]{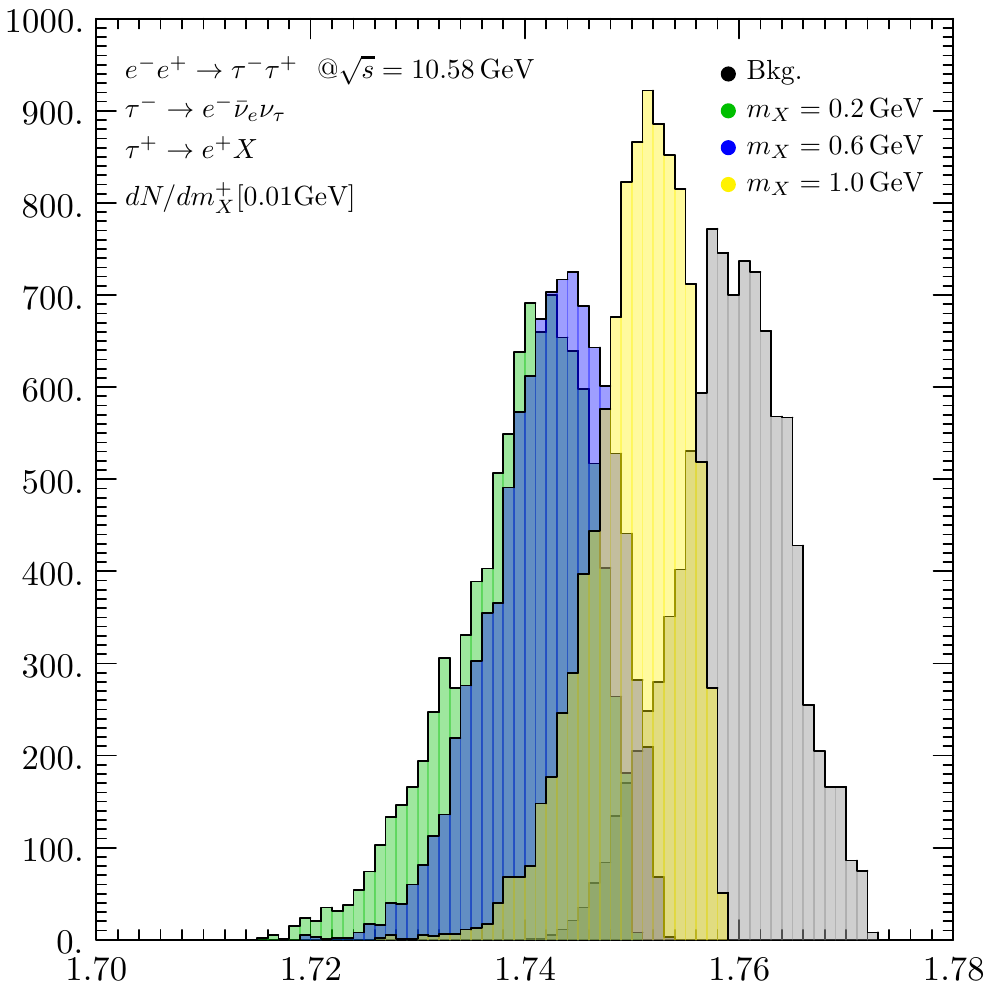}}
\hfill
\subfigure[]{\label{fig:MassX:Min}}
{\includegraphics[scale=0.578]{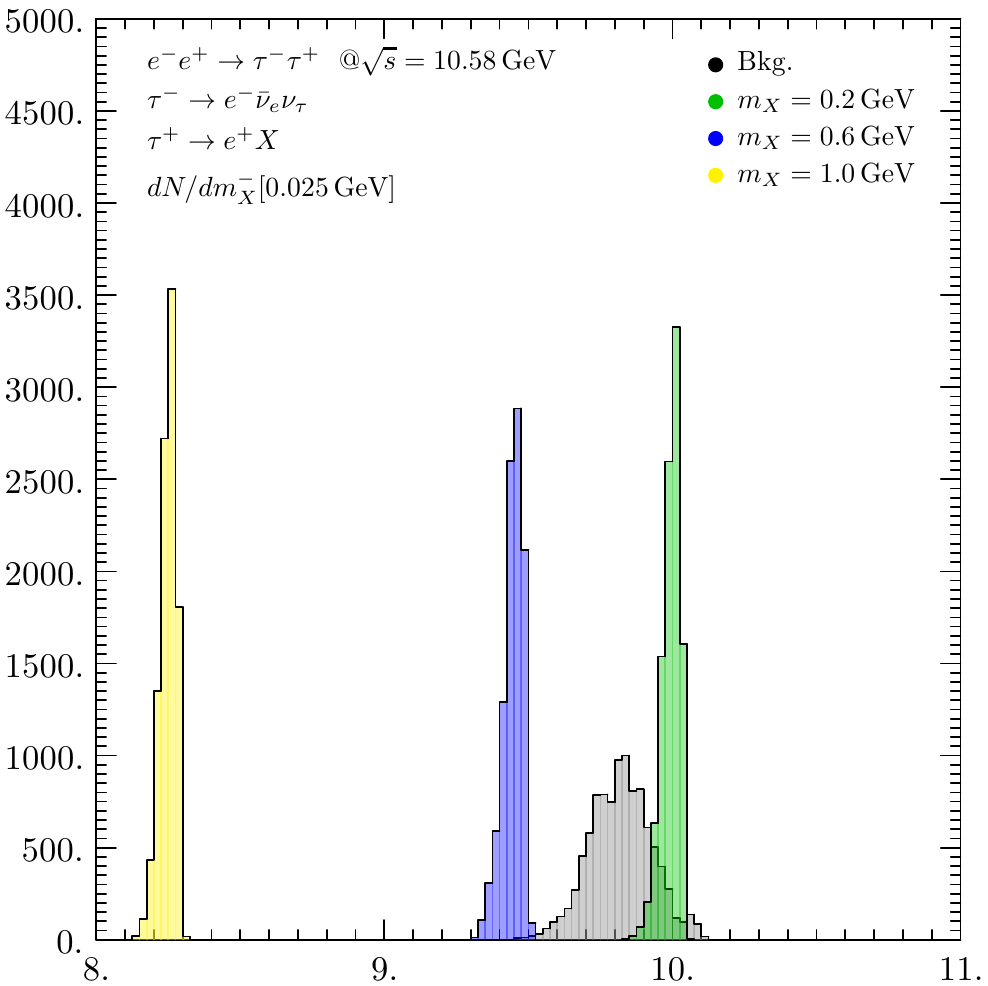}}
\hfill
\subfigure[]{\label{fig:MassX:Density}}
{\includegraphics[scale=0.168]{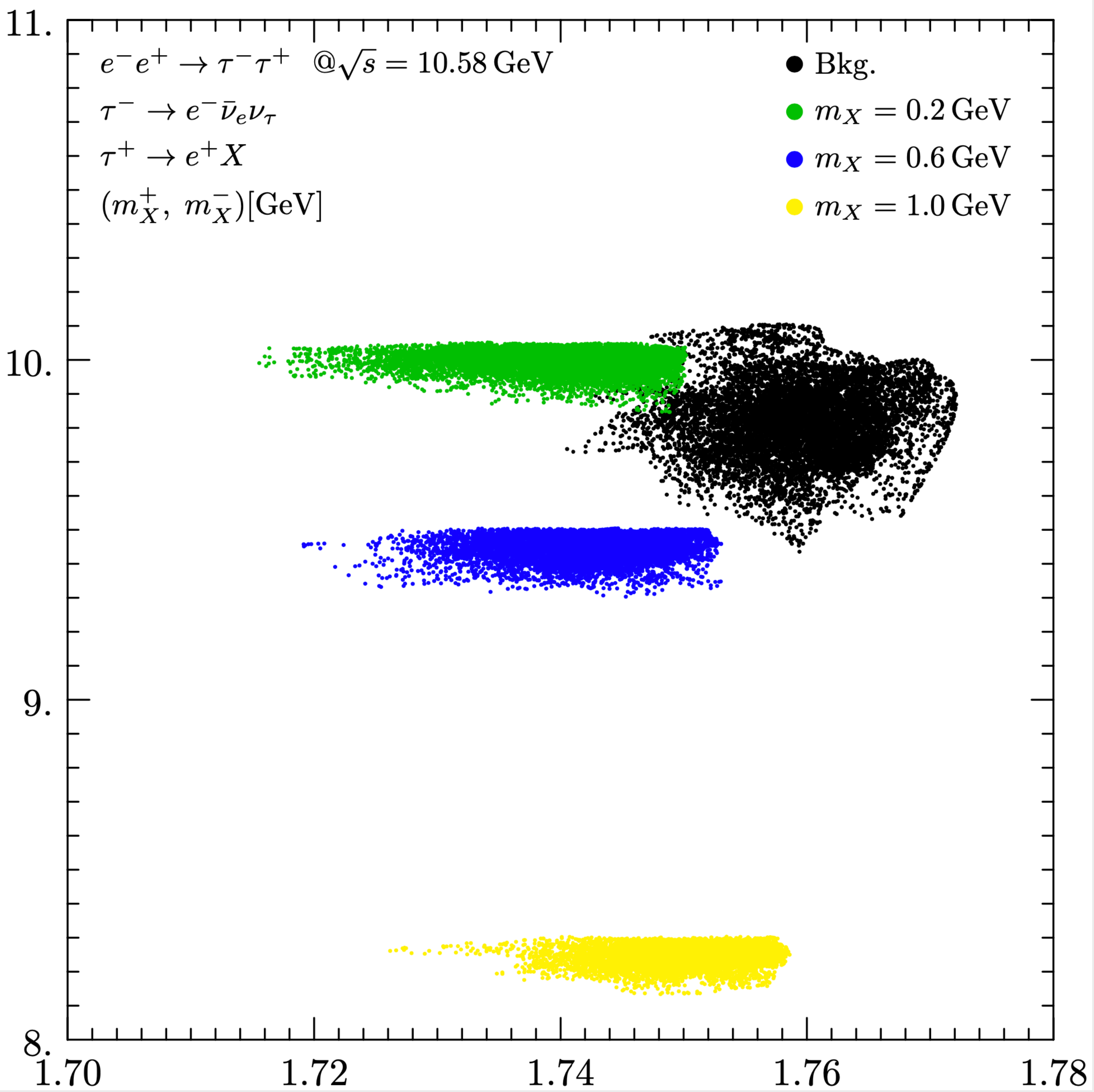}}
\caption{Fig.~\ref{fig:MassX:Max} and Fig.~\ref{fig:MassX:Min} are histograms of the 
observables $m_{D}^{+}$ and $m_{D}^{-}$, respectively. Fig.~\ref{fig:MassX:Density}
shows the scatter plots in the $(m_{D}^{+},\, m_{D}^{-})$ plane.
$10^{4}$ events are used for both the template of $\tau^{\pm}$ and the visible particle 
$e^{\pm}$. 
}
\label{fig:MassX}
\end{center}
\end{figure}
\end{widetext}
Fig.~\ref{fig:MassX:Max} and Fig.~\ref{fig:MassX:Min}  
are stacked histograms for $m_{D}^{+}$ and $m_{D}^{-}$. As expected $m_{D}^{-}$
is more sensitive to the mass of the invisible scalar, since the $\tau$-lepton has
a relatively large boost factor $\sim0.942$. Furthermore, while $m_{D}^{-}$ is
closing to the background when $m_{X}$ decrease, $m_{D}^{+}$ is leaving.
This implies a complementary property of our observables for measuring $m_{X}$.
Fig.~\ref{fig:MassX:Density} gives the density distributions in the $(m_{D}^{+}, m_{D}^{-})$
plane. One can see a clean linear correlation in the signal events, however a broad 
distributions in the background events. Fig.~\ref{fig:MassX:Mix} shows the distributions
for different significance $S/B$ with $100$ total number of events and $10^{4}$ events
for the templates of $\tau^{\pm}$. We can see that even for $S/B=1/9$ there is still
a possibility to observe the signal events. 

\textbf{\em{Example II: Lepton flavor violating decay of the Higgs}}.
Next we study a lepton flavor violating decay of the Higgs boson, 
$h\to\tau^{+}\mu^{-}$, which is predicted in many BSMs,  
such as supersymmetry~\cite{Arana-Catania:2013xma,Arhrib:2012ax}, 
models with flavor symmetries~\cite{Ishimori:2010au} 
or warped extra dimensions models~\cite{Perez:2008ee,Azatov:2009na,Albrecht:2009xr} 
and others~\cite{Goudelis:2011un,McKeen:2012av}.
The current upper limits on the branching ratios are $0.47\%$ and $ 0.28\%$~\cite{ATLAS:2019xlq,Aad:2019ugc}
for $h\to e\mu$ and $h\to \mu\tau$, respectively. 
Lepton flavor violating $Z$ boson decays are also 
predicted by BSMs~\cite{Illana:2000ic,GABBIANI1988398,PhysRevD.32.306},
and have been investigated by ATLAS~\cite{Aad:2020gkd}. The major backgrounds
are $h/Z\to\tau^{+}\tau^{-}$ and $h/Z\to\mu^{+}\mu^{-}$. In most of the 
investigated channels, the backgrounds is significantly larger than the 
signal~\cite{ATLAS:2019xlq,Aad:2019ugc}.
Furthermore, the situation become very worse when 
we consider the case that both $Z$ and $h$ decay in a lepton flavor violating way.
Therefore, it is very important to have an alternative way for probing 
lepton flavor violating decays in both $Z$ and $h$ channels. In our approach,
the observables can be defined by following equation,
\bee
m_{D}^{2}(\mu^{\pm}) = ( p_{h} - p_{\mu^{\pm}} )^{2} \,.
\ene
Fig.~\ref{fig:HTM:All}
is the scatter plot in the $(m_{D}^{+}, m_{D}^{-})$ plane for various channels, and
$10^{4}$ events for both $h/Z$ and $\mu^{\pm}$ are used.
The very narrow distributions of $m_{D}^{\pm}(\mu^{-})$ and $m_{D}^{\pm}(\mu^{+})$
for the $h\to\tau^{+}\mu^{-}$ and $h\to\mu^{+}(\tau^{+})\mu^{-}(\tau^{-})$
channels are amplified in Fig.~\ref{fig:HTM:left} and 
Fig.~\ref{fig:HTM:Right}, respectively. We can clearly see that signal and backgrounds
are separated well, except for $m_{D}^{\pm}(\mu^{+})$ for the channels 
$h\to\tau^{+}\mu^{-}$ and $h\to\tau^{+}\tau^{-}$. However, this degenerate can be
completed lifted by investigating correlations between $m_{D}^{\pm}(\mu^{+})$ and
$m_{D}^{\pm}(\mu^{-})$. For the background decay channel $h\to\tau^{+}\tau^{-}$,
$m_{D}^{\pm}(\mu^{+})$ and $m_{D}^{\pm}(\mu^{-})$ distributions are always 
represented by the purple region. On the other hand, 
$m_{D}^{\pm}(\mu^{+})$ and $m_{D}^{\pm}(\mu^{-})$ distributions for
the signal decay channel $h\to\tau^{+}\mu^{-}$ are represented by
green and black regions. Therefore, the signal events can be distinguished by
the distribution of $m_{D}^{\pm}(\mu^{-})$. Similar property is expected for possible
signal decay process $Z\to\tau^{+}\mu^{-}$. 
\begin{widetext} 
\mbox{}
\begin{figure}[thb]
\begin{center}
\subfigure[]{\label{fig:HTM:All}}
{\includegraphics[scale=0.178]{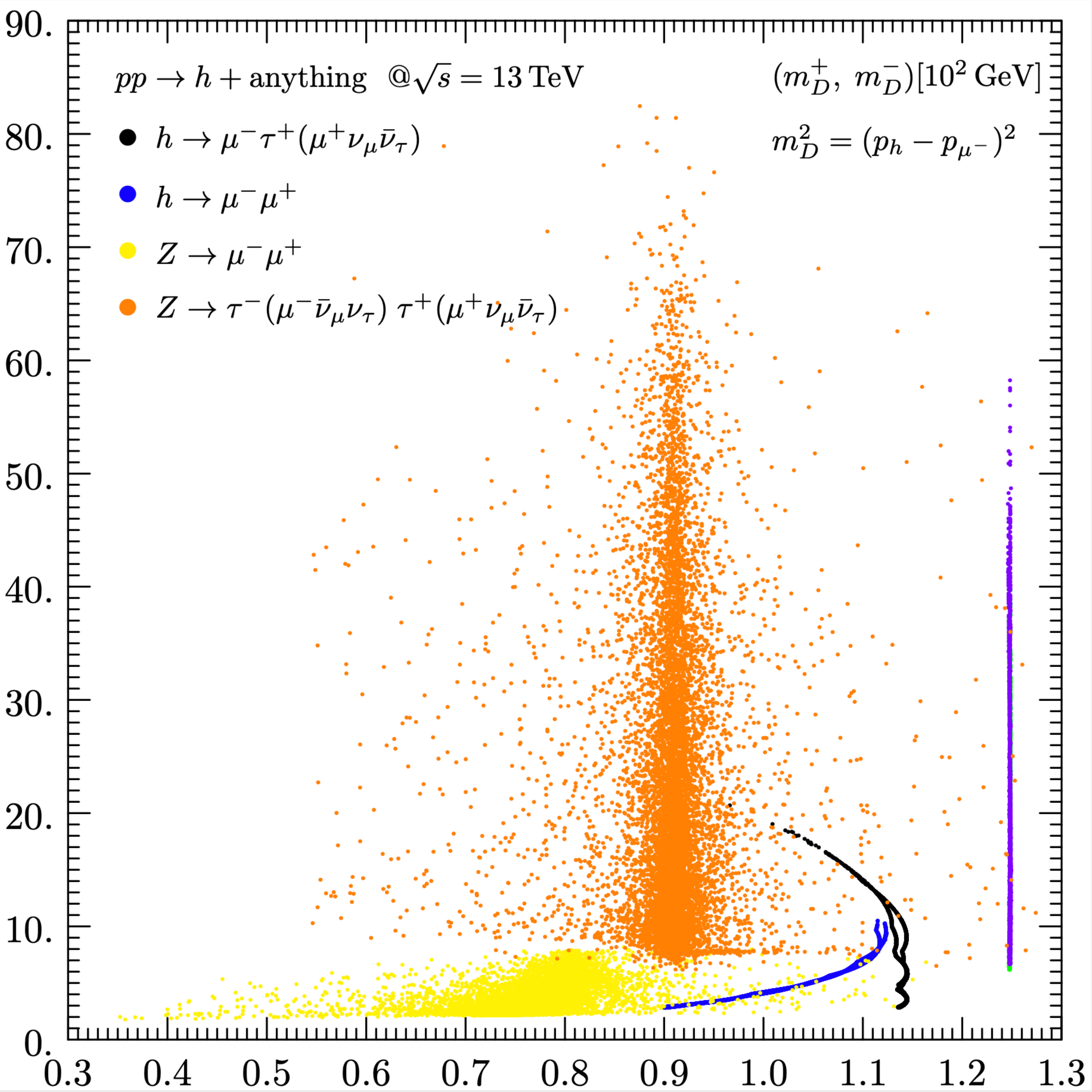}}
\hfill
\subfigure[]{\label{fig:HTM:left}}
{\includegraphics[scale=0.175]{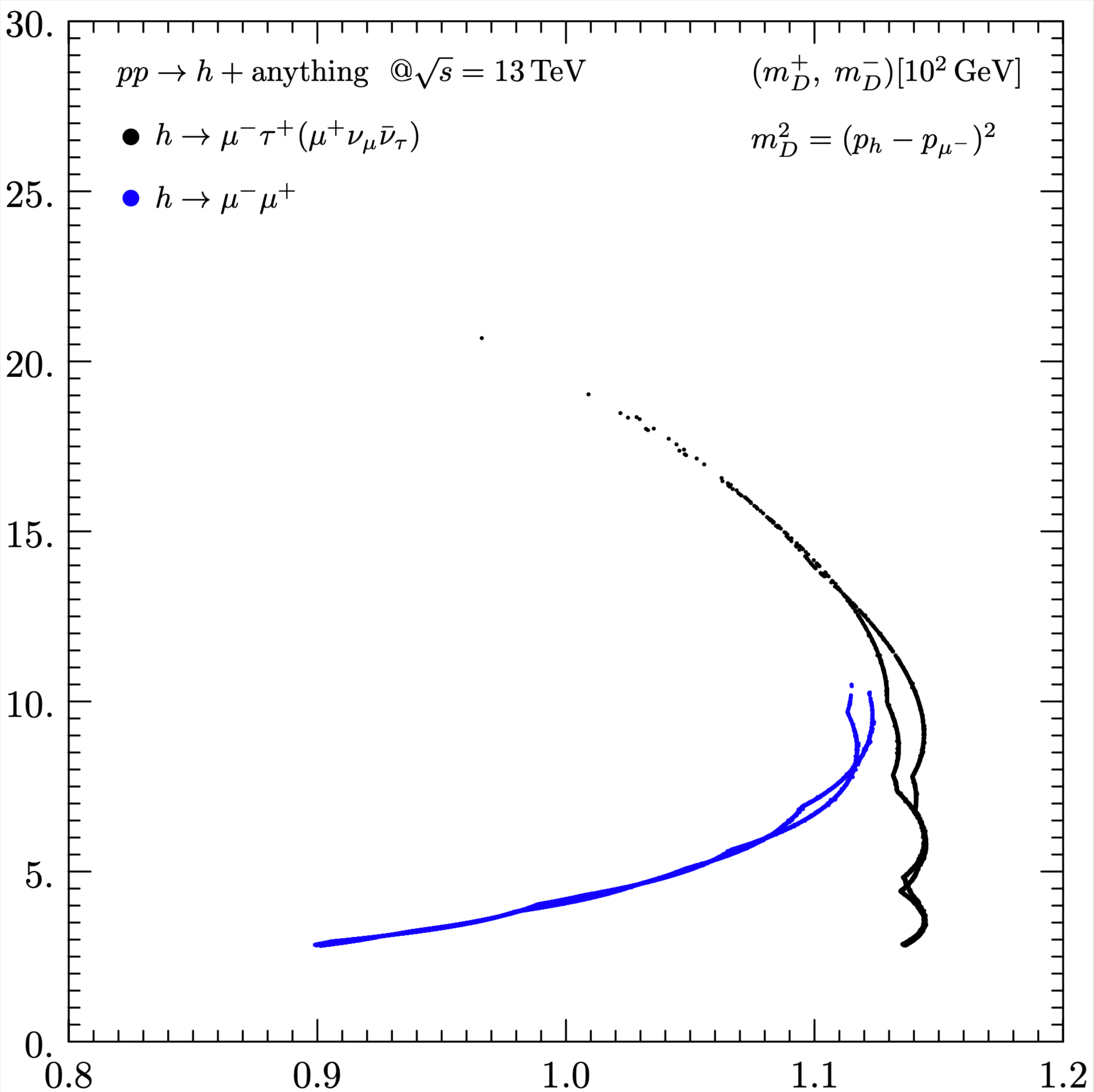}}
\hfill
\subfigure[]{\label{fig:HTM:Right}}
{\includegraphics[scale=0.178]{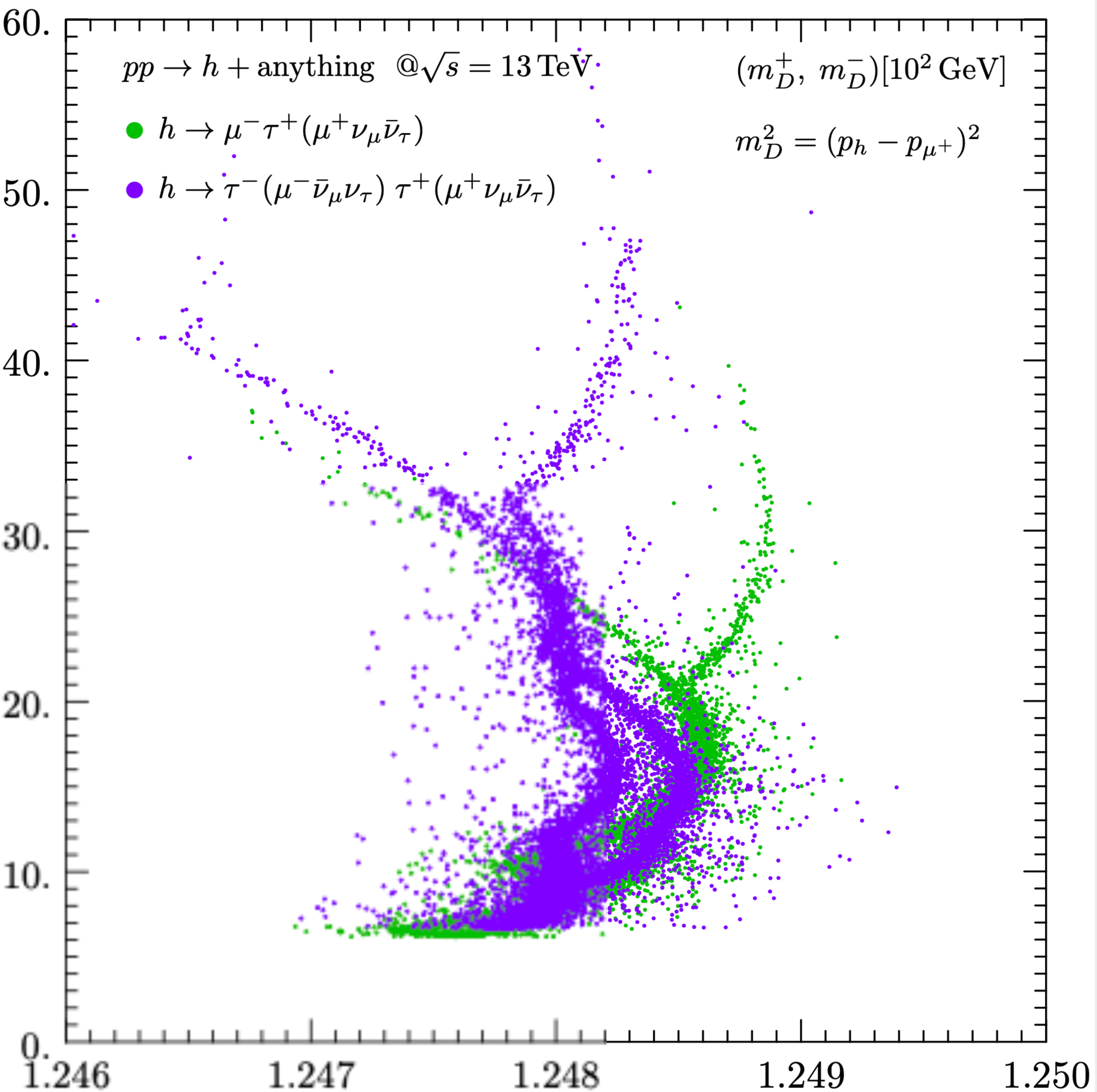}}
\caption{Scatter plots for decay channels $h/Z\to (\mu^{+}/\tau^{+})(\mu^{-}/\tau^{-})$.
$10^{4}$ events are used for both the template of $h/Z$ and the visible particle $\mu^{\pm}$.
Fig.~\ref{fig:HTM:left} and Fig.~\ref{fig:HTM:Right} are amplified representation of
the very narrow distributions regions in Fig.~\ref{fig:HTM:All}. 
}
\label{fig:MassX}
\end{center}
\end{figure}
\end{widetext}

\begin{figure}[H]
\begin{center}
\includegraphics[scale=0.205]{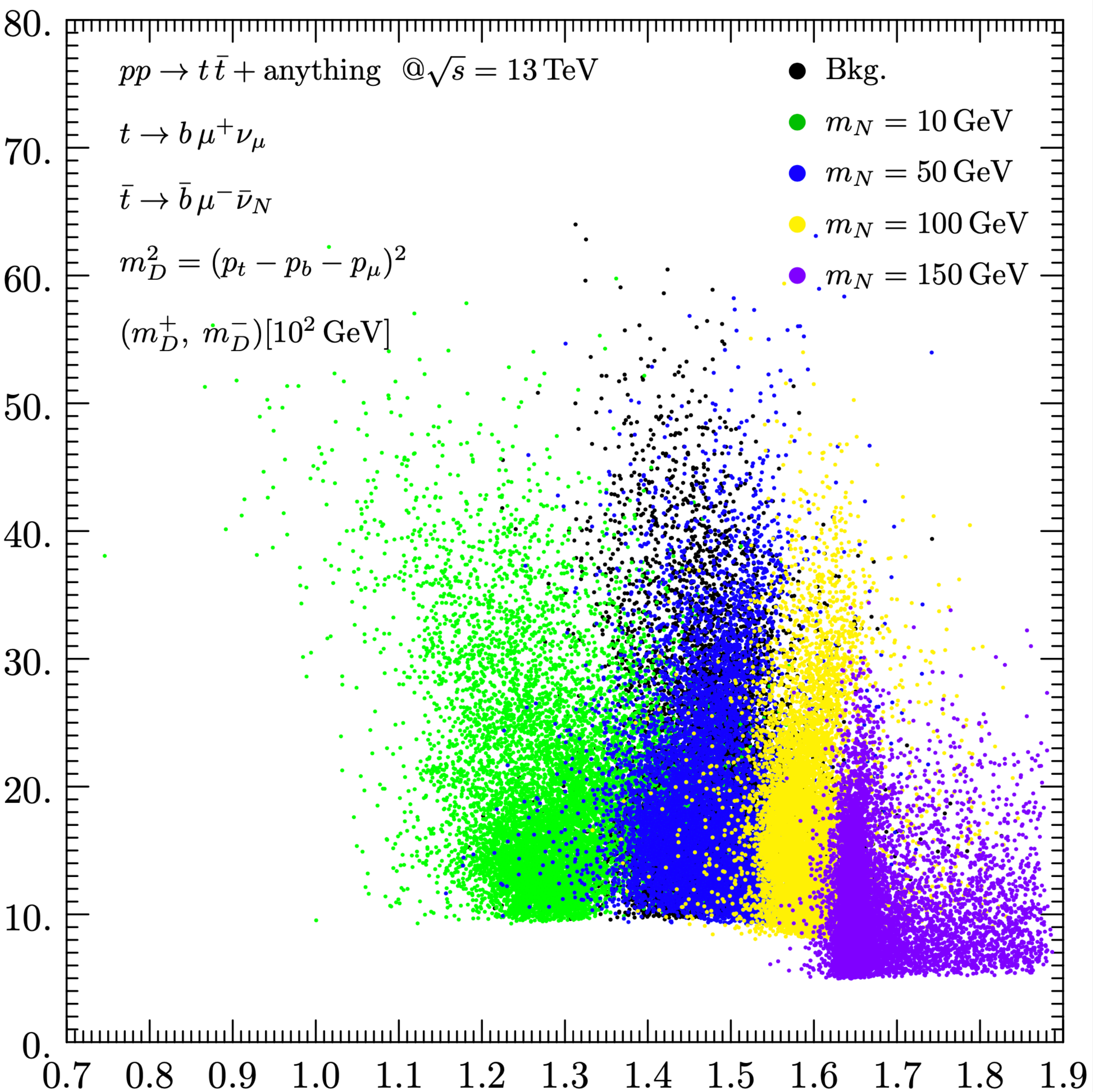}
\caption{Scatter plots for decay channels $\bar{t} \to \bar{b} + \mu^{-} + \bar{\nu}_{N}$
and $ t \to b + \mu^{+} + \bar{\nu}_{\mu}$.
$10^{4}$ events are used for both the template of $t/\bar{t}$ and the 
visible particles $b/\bar{b}$ and $\mu^{\pm}$.
}
\label{fig:nTopMuon:Density}
\end{center}
\end{figure}

\textbf{\em{Heavy neutrino}}. Next we give an example in which the invisible 
particle appears in a second subsequent decay. we illustrate this example by
introducing a heavy neutrino which is necessary in the well-known seesaw 
mechanism. Here we consider a relatively light 
neutrino~\cite{He:2009ua,Adhikari:2010yt,Boucenna:2014zba} 
emerging from decay of a top-quark~\cite{BarShalom:2006bv,Si:2008jd,Liu:2019qfa},
\bee
\bar{t} \to \bar{b} + W^{-}, \;\;W^{+} \to \mu^{-} + \bar{\nu}_{N}\,.
\ene
Furthermore, its interaction with the $W$-boson is assumed to be 
right-handed~\cite{Alcaide:2019pnf}.
Off-shell effect of $W^{-}$ when $m_{N} > m_{W}$ is also taken into account.  
Observables are defined as the boundary values of $m_{D}^{2}$ defined as
\bee
m_{D}^{2} = (p_{t} - p_{b} - p_{\bar{\ell}})^{2}\,.
\ene
Fig.~\ref{fig:nTopMuon:Density} shows the distributions of $m_{D}^{\pm}$ for
different mass of $\nu_{N}$. Except for the case $m_{N}=50\gev$, distributions
of $m_{D}^{\pm}$ are separated from the background from 
$\bar{t} \to \bar{b} + \mu^{-} + \bar{\nu}_{\mu}$. Again, degenerate between 
backgrounds and signals with $m_{N}$ around $50\gev$ is expected to be lifted 
by studying alternative definitions of $m_{D}^{2}$. For instances 
$m_{D}^{2}= (p_{t} - p_{\bar{\ell}})^{2}$ or 
$m_{D}^{2} = (p_{t} - p_{b})^{2}$ in case of that $m_{N} > m_{W}$.
On the other hand, since the top-quark pair can be produced in both
gluon and quark channels, it is also possible to separate signals and 
backgrounds by studying $m_{D}^{\pm}$ channel by channel.
We leave those studies in a future work.
\vspace{-0.2cm}

In summary we provided a general approach for studying semi-invisible decay
process. The observables $m_{D}^{\pm}$ are complementary in the sense they
are sensitive to production kinematics in different regions, and most importantly
are useful in case of that the significance $S/B$ is relatively low.

\section*{Acknowledgements}
This study is supported by the National Natural Science Foundation of China under Grant No. 11705113, and Natural Science Basic Research Plan in Shaanxi Province of China under Grant No. 2018JQ1018, and the Scientific Research Program Funded by Shaanxi Provincial Education Department under Grant No. 18JK0153.

\bibliography{invisTau}

\end{document}